\documentclass[prd,twocolumn,nofootinbib,superscriptaddress]{revtex4-2}

\usepackage{graphicx}
\usepackage{dcolumn}
\usepackage{bm}
\usepackage{amsmath, amssymb, mathrsfs}
\usepackage{epsfig}
\usepackage[english]{babel}
\usepackage[normal]{subfigure}
\usepackage{multirow}
\usepackage{xcolor}
\usepackage{hyperref}

\begin{document}

\title{Enhancement of Gravitationally Induced Entanglement via Quantum Squeezing}

\author{Sijo K. Joseph}
\email{skizhakk@gitam.edu}
\affiliation{Theoretical and Computational Physics Research Unit, Department of Physics, GITAM Deemed to be University, Bengaluru, Karnataka, India - 562163}

\date{\today}

\begin{abstract}
Gravitational generation of quantum entanglement on squeezed coherent states in a system of two quantum mechanical particles is explored. The particles are modeled as harmonic oscillators, and their gravitational interaction is described by an interaction potential. Dynamics of the gravitationally induced entanglement is explored using the linear entropy of entanglement. An analytical expression for the linear entropy is obtained via Gaussian integration and covariance method. In the short-time weakly coupled regime, it is observed that the entanglement is proportional to the square of the time interval, square of the interaction strength, and proportional to the squeezing parameters of the modes. These results provide insights into the effects of quantum squeezing on entanglement generation in systems with gravitational interactions. Analytically and numerically it is shown that momentum squeezing can enhance gravitationally induced quantum entanglement.
\end{abstract}

\maketitle

\section{Introduction} \label{intro}
Continuous variable quantum entanglement dynamics is of significant interest in quantum information and computation~\cite{nielsen00,Braunstein05,Andersen10,Weedbrook12}. Quantum entanglement, which describes non-classical correlations between particles, is crucial for various quantum protocols~\cite{RPMK_Horodecki_09, Ursin07, Yin2020, Krelina2021}. 
One of the challenges in theoretical physics is to establish a proper theory of quantum gravity. Recent explorations has opened, several table-top experimental proposals to test the quantum entanglement generated via the  gravitational interaction of the small particles~\cite{Feifan23,Carlesso_2019, Krishnada2020,PRXCarney21,Tilly21,Schut22,Bahrami2015}. Once we detect such a gravitationally induced entanglement, it will reveal interesting features of gravity. Still it is debatable such an entanglement can ensure the quantum or classical nature of gravity~\cite{Kemal22,Kemal23}. Using quantum information theory arguments, it is argued that the detection of gravitationally induced entanglement will indicate the quantum features of gravity. In quantum information theory perspective, local operation with classical communication cannot generate entanglement in an initially unentangled state. Entangled harmonics oscillators and interesting system to analyze, and it is already known that an opto-mechanical system with optical cavity and mirror can be modeled as a quantum harmonic oscillator in a sufficiently smaller temperature~\cite{Aspelmeyer14, Regal_2011, sarma2022}. Detecting these entanglement will be extremely sensitive, first we need to preserve the quantum state of the particles, secondly gravitational effects will be much weaker which results into a small entanglement generation. Hence it is worthwhile to look into ways to improve this effect using proper quantum mechanical states, quantum squeezing is a nice quantum technique to achieve that goal.

In quantum optics community, quantum squeezing has been identified as a potential tool to control quantum entanglement~\cite{Esteve08, Alebachew07,icfo_squeeze,Er13}. Previous works had explored entanglement enhancement through squeezing in different systems, including the Jaynes-Cummings model and systems of coupled harmonic oscillators~\cite{Furuichi01}. The effectiveness of initial squeezing on entanglement enhancement and its relationship with chaotic dynamics is investigated in many research works~\cite{Wang_Sanders,Zhang08,Chung07,skj_physlett,skj_epjd}.  It is well known in quantum chaos and quantum optics community that the quantum squeezing can play a significant role in entanglement enhancement~\cite{Er13,Xiang09,skj_physlett,skj_epjd}. Notably, entanglement was found to persist even in a decohering environment with high temperature when the normal modes are squeezed \cite{Galve10}.  In this study, the gravitational generation of entanglement on squeezed coherent states is analyzed using a system of two quantum mechanical particles modeled as harmonic oscillators. These two harmonics oscillators are subjected to a gravitational interaction potential  generated by the particle mass itself. We investigate the short-time dynamics of entanglement by considering the linear entropy of entanglement. The main goal of this study is to analyze how the squeezing parameters of the initial squeezed coherent states  affect the gravitationally induced entanglement dynamics.

Our manuscript is organized as follows. In Section II, we describe the model with the gravitational interaction potential and its quantization. In Section III, a brief overview of the squeezed coherent state in the continuous position basis is given. In Section IV, an analytical expression for the entanglement dynamics of the time-evolved quantum state is derived by evaluating the linear entropy of entanglement via the Gaussian integral method. In Section V, we analyze the result in the weak-interaction regime. In Section VI, the same result is derived independently using the covariance matrix method. In Section VII, the exact entanglement dynamics is explored numerically with the split-operator Fourier method, and the validity of our analytical results is examined in the long-time regime. Finally, in Section VIII, we summarize our findings.
\section{Model description}
We are considering two quantum mechanical particles separated by a distance $L$ as shown in Fig.1.
We are treating the particles as harmonic oscillators, the gravitational interaction between them give rise to an interaction Hamiltonian $H_{int}$. We consider a general two-dimensional classical Hamiltonian of the form
\begin{equation}\label{GenHam1}
H=H_{1}+H_{2}+H_{int},
 \end{equation}
where $H_{1}$, $H_{2}$ is the Hamiltonian of two identical particles considered as harmonic oscillators and they are interacting via an interaction Hamiltonian $H_{inter}$. In this specific case, we consider that the particle interacts with its own gravitational field.  In terms of the position coordinates $x$ and $y$,  Hamitonian can be written as,
\begin{equation}\label{GenHam2}
H=\frac{1}{2M}(p_{x}^{2}+p_{y}^{2})+\frac{1}{2}K\,(x^{2}+y^{2})-\frac{G\,M^2}{L+(y-x)}.
 \end{equation}
When the separation between the oscillators are very high i.e $L>>(y-x)$, the series expansion of the potential will lead to the following Hamiltonian. 
\begin{figure} 
  \centering
  \includegraphics[width=1.0\linewidth]{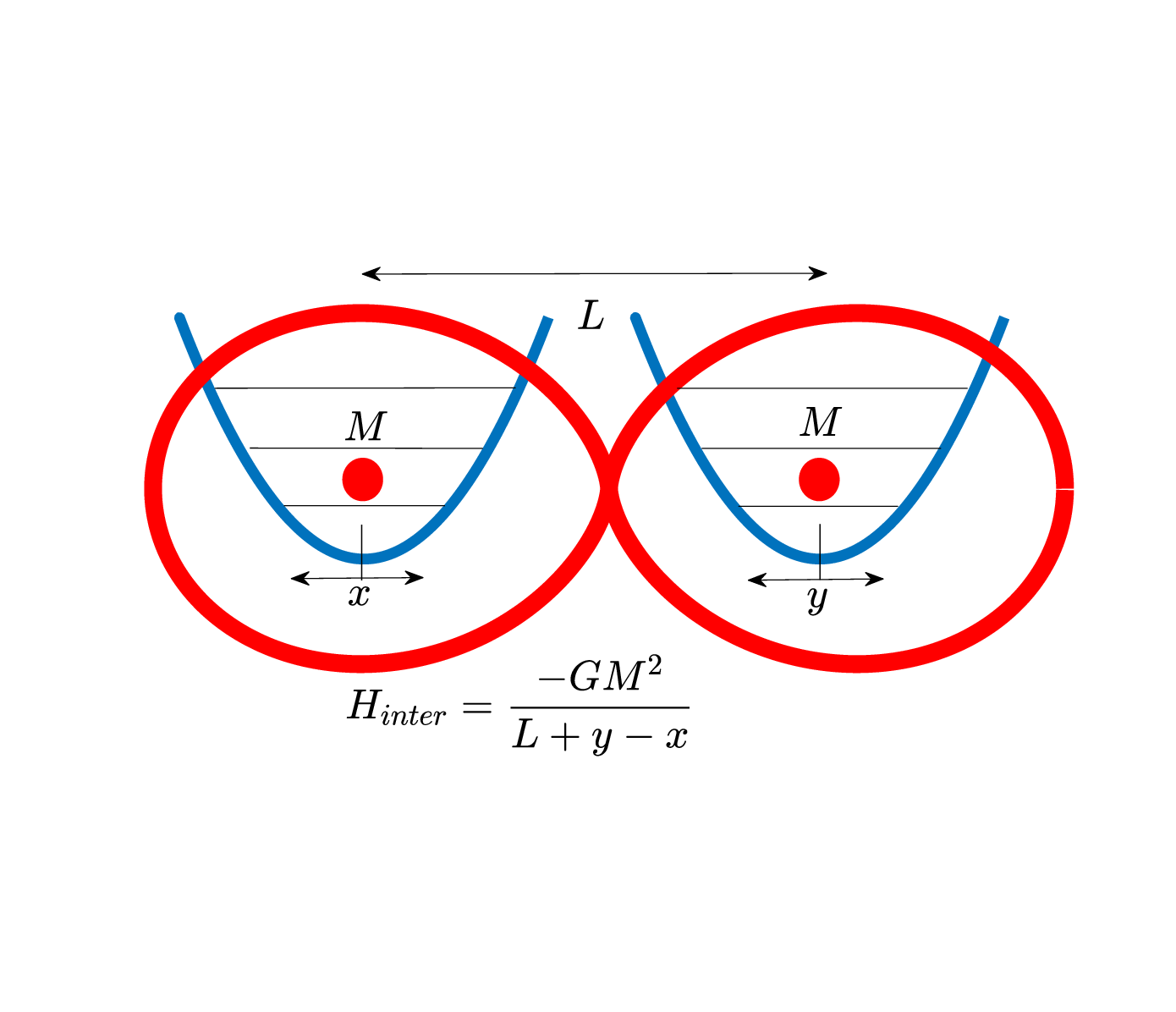}
  \caption{Figure shows two identical quantum particle of mass $M$, trapped in a harmonic potential separated by a distance $(L)$. These particles interact via its own gravitational field. Due this gravitational interaction ($H_{int}$), particle's quantum state can get entangled each other.}
  \label{fig:1}
\end{figure}

%

\begin{eqnarray}\label{GenHam3}
H=&\frac{1}{2M}(p_{x}^{2}+p_{y}^{2})+\frac{K}{2}(x^{2}+y^{2})\nonumber\\
     & -\gamma \sum_{n} \frac{1}{L^n} \sum_{r} (-1)^r {n\choose{r}} {x}^{n-r}{y}^{r}
 \end{eqnarray}
where $\beta=\frac{GM^2}{L} $.
 One can wrtie $H=H_{0}+V_{g}$, where $\hat{H_{0}}= \frac{1}{2M}(p_{x}^{2}+p_{y}^{2})$ is the two-dimensional harmonic oscillator Hamiltonian and  $\hat{V_{g}}=\gamma \sum_{n}  \frac{1}{L^n}  \sum_{r} (-1)^r {n\choose{r}} {x}^{r}{y}^{n-r}$  is the gravitational interaction part expanded as a power series. One can also perceive this problem as a particle moving in a two dimensional potential  $V(x,y)=\frac{1}{2}(x^2+y^2)+\frac{\gamma}{L+y-x}$.
By looking into the gravitationally induced interaction term, it can be seen that every external perturbation is of the form $x^{m}y^{n}$. The terms which will contribute to the entanglement of the initial product state involves $x^{m}y^{n}$ where $m>0$ and $n>0$. If $m=0$ or $n=0$ in any of the terms, then it won't produce any effect on bipartite entanglement since it is no longer an interaction term. One can expand the gravitational interaction up to the order of $({(x-y)}/{L})^2$ and ignore the higher order terms, then the following Hamiltonian is obtained,
\begin{eqnarray}\label{GenHam4}
H&=&\frac{1}{2M}(p_{x}^{2}+p_{y}^{2})+\frac{(K-\frac{2\gamma}{L})}{2}\Big((x-\frac{\gamma}{2L})^{2}+(y+\frac{\gamma}{2L})^{2}\Big)\nonumber\\
   &  & +\frac{2\gamma}{L^2} x y -\frac{2\gamma^2}{4L^2}-\gamma.
 \end{eqnarray}
Simply making a change of variable $\bar{K}={(K-2\gamma/{L})}$, $\bar{x}=x-{\gamma}/{(2L)}$  and $\bar{y}=y+{\gamma}/{(2L)}$, then the Hamiltonian get simplified to,

\begin{eqnarray}\label{GenHam42}
H&=&\frac{1}{2M}(p_{\bar{x}}^{2}+p_{\bar{y}}^{2})+\frac{\bar{K}}{2}(\bar{x}^{2}+\bar{y}^{2})\nonumber\\
   &  & +\frac{2\gamma}{L^2}(\bar{x}+\frac{\gamma}{2L})(\bar{y}-\frac{\gamma}{2L}) -\frac{2\gamma^2}{4L^2}-\gamma.
 \end{eqnarray}
Ignoring the constant shift in energy $(-\frac{2\gamma^2}{4L^2}-\gamma)$ as well as the non-interacting terms (which doesn't contribute to entanglement), one can obtain,
\begin{eqnarray}\label{GenHam43}
H=\frac{1}{2M}(p_{\bar{x}}^{2}+p_{\bar{y}}^{2})+\frac{\bar{K}}{2}(\bar{x}^{2}+\bar{y}^{2})
     +\lambda\,\bar{x}\,\bar{y},
 \end{eqnarray}
where $\lambda=2\gamma/L^2=2GM^2/L^3$.
Here the problem is reduced to the two-dimensional harmonic oscillator problem with an interaction potential $\lambda \bar{x} \bar{y}$ and this interaction term is responsible for the gravitationally induced quantum entanglement. 

Proceeding with the quantization of the problem, the Schr\"{o}dinger equation can be written as,
\begin{equation}\label{schrodi2}
i\hbar\frac{\partial}{\partial t}\psi({x,y,t})  =  (\hat{H_{0}} + \hat{V_{g}}) \psi({x,y,t}),
\end{equation}
where $\hat{H_{0}}= \frac{-\hbar^2}{2m}\nabla^2+\frac{K}{2}(x^{2}+y^{2})$ is the two-dimensional Harmonic Oscillator Hamiltonian and $\hat{V_{g}}=\lambda x y$
is the gravitational interaction part. 
The time evolution of the quantum state $\psi({x,y,t})$ is given by
\begin{equation}\label{evol_eq}
\psi({x,y,t})=\hat{U}(t)\psi({x,y,0}),
\end{equation}
where the time evolution operator is given by
\[
\hat{U}(t)= \exp{\left({\frac{-i t \mkern1mu} {\hbar}}(\hat{H_{0}} + \hat{V_{g}})\right)} \,.
\]

In this work, a pure continuous variable bipartite state is considered, which is the tensor product of two squeezed coherent states. The reduced density function of the first subsystem is obtained by integrating over the second subsystem. To quantify continuous variable entanglement, the linear entropy of entanglement is used. 

First, we write a pure continuous bipartite state as follow:
\begin{equation}
{|\psi\rangle}_{12}=\int {\psi(x,y)|x\rangle |y\rangle dx\;dy},
\end{equation}
where $|x \rangle$  and $|y\rangle$ are the continuous basis representation of the position operators of the first and second particle, respectively. 
One can find the density matrix of this bipartite quantum state as follows,
\begin{eqnarray}
\hat{\rho}_{12}&=&|\psi_{12}\rangle \langle \psi_{12}| \nonumber\\
&=&\int {\psi(x,y)\psi^{*}(x',y')|x\rangle |y\rangle \langle x'| \langle y'| dx\;dy\;dx'\;dy'\;} \nonumber \\
\end{eqnarray}

The reduced density function of the first subsystem $\hat{\rho_{1}}$ can be obtained by integrating over the one of the particle mode i.e $\hat{\rho_{1}}=\mbox{Tr}_{2}(\hat{\rho}_{12})$.
In the continuous basis  the reduced density matrix can be expressed in terms of the bipartite wave function $\psi({x,y})$, i.e,

\begin{equation}\label{ReduDen}
\rho_{1}({x,x'})=\int {\psi({x,y}){\psi}^{*}({x',y}) dy\;},
\end{equation}
Trace of the square of the reduced density matrix can be written as a double integral,
\begin{equation}\label{TrSqeq}
\mbox{Tr}(\hat{\rho_{1}}^2)= \int\int {\rho_{1}({x,z}){\rho_{1}} ({z,x}) dx\;dz\;}.
\end{equation}
Writing the  linear entropy ($S_{L}(t)=1-\mbox{Tr}(\hat{\rho_{1}}^2)$) in terms of the continuous basis representation as follows,
\begin{equation}\label{tr_dub_intgrl}
S_{L}=1-\int\int {\rho_{1}({x,z}){\rho_{1}} ({z,x}) dx\;dz\;}.
\end{equation}

\section{The squeezed coherent state}
The squeezed coherent state can be generated via  displacement and squeezing operators acting on the vacuum state\cite{Caves81,JagdishRai88,FanHong89,Moller96}. The displacement operator is given by

\begin{equation}
\hat{D}(\alpha_{k}) =  \exp{({{\alpha}_{k}} \hat{a_{k}}^{\dagger}-{\alpha}_{k}^* \hat{a_{k}})} \,
\end{equation}
and
\begin{equation}
\hat{S}(\zeta_{k}) =  \exp{(\frac{1}{2}{\zeta}_{k} \hat{a_{k}}^{\dagger^{2}}-\frac{1}{2}{{\zeta}_{k}}^* \hat{a_{k}}^{2})} \,.
\end{equation}
The squeezed coherent state is thus defined as
\begin{eqnarray}
|\alpha_{k},\zeta_{k}\rangle & =  &\hat{D}({\alpha}_{k}) \hat{S}({\zeta}_{k}) |0 \rangle \,.
\end{eqnarray}
Here $\alpha_{k}=|\alpha_{k}|e^{ \mathrm{i} \phi_{k}}$ and $\zeta_{k}=r_{k}\,e^{ \mathrm{i} \theta_{k}}$ are complex numbers. $\alpha_{k}$ is related to the phase space variables $(q_{k},p_{k})$, which is given by
\begin{equation}
\alpha_{k}=\frac{1}{\sqrt{2\hbar}}(q_{k}+{i}p_{k}),
\end{equation}
where $k=1,2,$  respectively.
According to M{\o}ller, J{\o}rgensen and Dahl, the squeezed coherent state in the position basis $\langle x|\alpha_{k},\zeta_{k}\rangle =\psi(x,\alpha_{k},\zeta_{k}) $ can be written as
\begin{eqnarray}
&\psi(x,\alpha_{k},\zeta_{k}) =  {\left(\frac{M\omega}{\pi\hbar}\right)}^{1/4}{(\cosh{{r}_{k}}+{\mathrm{e}^{\mathrm{i} \theta}}\sinh{{r}_{k}})}^{-1/2} \nonumber \\ 
&\hspace{-2em} \exp{\left\{-\frac{M\omega_{k}}{2\hbar}\left(\frac{\cosh{{r}_{k}}-{\mathrm{e}^{\mathrm{i} \theta}}\sinh{{r}_{k}}}{\cosh{{r}_{k}}+{\mathrm{e}^{\mathrm{i} \theta}}\sinh{{r}_{k}}}\right) {(x-q_{k})}^{2}+\frac{i}{\hbar}{p_{k}(x-{q_{k}/2)}} \right\}}.\nonumber \\
\label{sqstatepos}
\end{eqnarray}
We are taking an intial state which is a product state of two squeezed coherent state i.e, $|\alpha_{1},\zeta_{1}\rangle \otimes |\alpha_{2},\zeta_{2}
\rangle $. 
In the continous position basis, it is just a product of two wavefuntions ie $\psi(x,y,\alpha_{1},\zeta_{1},\alpha_{2},\zeta_{2}) =\psi(x,\alpha_{1},\zeta_{1})
\psi(y,\alpha_{2},\zeta_{2}) $. 
This is just the projection of $|\alpha_{1},\zeta_{1}\rangle \otimes |\alpha_{2},\zeta_{2}\rangle $ in the $x,y$ continuous basis i.e. $\langle x|\alpha_{1},
\zeta_{1}\rangle \otimes \langle y|\alpha_{2},\zeta_{2}\rangle$. We would like to express Eq.~\ref{sqstatepos} in a more compact manner once the squeezed coherent state is represented in a shifted coordinate $\bar{x}=(x-x_{k})$.
\begin{equation}\label{gaussx}
\psi(\bar{x})=  {\left(\frac{\Re{(A_{1})}}{\pi}\right)}^{1/4}\exp{\left(-\frac{1}{2}({A}_{1}{\bar{x}}^{2}+iB_{1}\bar{x}+iC_{1})\right)}
\end{equation}
Here $A_{1}$ is a complex number while $B_{1}$ and $C_{1}$ are real numbers. Similarly for the second particle, the wavefunction can be writen in a compact manner,
\begin{equation}\label{gaussy}
\psi(\bar{y})=  {\left(\frac{\Re{(A_{2})}}{\pi}\right)}^{1/4}\exp{\left(-\frac{1}{2}({A}_{2}{\bar{y}}^{2}+iB_{2}\bar{y}+iC_{2})\right)}.
\end{equation}
The coefficients are given by $A_{k}=\frac{M\omega}{\hbar}\left(\frac{\cosh{{r}_{k}}-{\mathrm{e}^{\mathrm{i} \theta}}\sinh{{r}_{k}}}{\cosh{{r}_{k}}+{\mathrm{e}^{\mathrm{i} \theta}}\sinh{{r}_{k}}}\right) $, $B_{k}={p_{k}}/{\hbar}$, $C_{k}={p_{k}q_{k}}/{\hbar}$, $\bar{x}=(x-x_{k})$ and $\bar{y}=(y-y_{k})$. Now the notation for the tensor product state gets simplified  and the initial state is denoted by $\psi(\bar{x},\bar{y},0)$.  Here $\Re{(A_{k})}$ represents the real part of the coefficient $A_{k}$.
Consider the squeezed coherent wave function associated with the ${x}$ variable in the position basis $\psi({x},\alpha_{1},\zeta_{1})$, as given in Eq. \ref{sqstatepos}. Here, $\alpha_{1}$ is the center of the Gaussian wave packet and $\zeta_{1}$ is the squeezing parameter. Similarly, we consider the squeezed coherent wave function associated with the ${y}$ variable in the position basis. 
Now, we take the tensor product of the squeezed coherent state (Eq.~\ref{gaussx} and Eq.~\ref{gaussy}) and the initial wave packet at time $t=0$ is given by,
\begin{eqnarray}
\psi(\bar{x},\bar{y})=  {\left(\frac{\Re{(A_{1})}}{\pi}\right)}^{1/4} {\left(\frac{\Re{(A_{2})}}{\pi}\right)}^{1/4}\nonumber \\ \exp{\left(-\frac{1}{2}({A}_{1}{\bar{x}}^{2}+iB_{1}\bar{x}+iC_{1})\right)} \nonumber \\ \exp{\left(-\frac{1}{2}({A}_{2}{\bar{y}}^{2}+iB_{2}\bar{y}+iC_{2})\right)}.\label{2d_sq_vaccum}
\end{eqnarray}

The time evolution of the wave packet is then calculated by substituting Eqs.~\ref{Zassenhaus} and  \ref{2d_sq_vaccum} into Eq.~\ref{evol_eq}.  Once we evolve the state using the unitary operator described by the Hamiltonian, both the particle gets an entangled state i.e $\psi(\bar{x},\bar{y},\Delta t)$.  Once can systematically compute the entanglement entropy of  $\psi(\bar{x},\bar{y},\Delta t)$, by analyzing the linear entropy of entanglement using Eq.~\ref{tr_dub_intgrl}, which is just a matter of computing Gaussian Integrals.
\section{Gravitationally induced entanglement dynamics}
The short time evolution of the wave function is obtained through the evaluation of Zassenhaus formula\cite{Zassenhaus_Ref}, followed by truncating all the higher order correction terms in $t$, with the time evolution operator expressed in the following way:
\begin{equation}\label{Zassenhaus}
\hat{U}(t) = \exp({\frac{-i \Delta t \mkern1mu} {\hbar}}\hat{H})\approx   \exp({\frac{-i \Delta t \mkern1mu} {\hbar}}\hat{V_{\lambda}}) \exp({\frac{-i \Delta t \mkern1mu} {\hbar}}\hat{H_{0}}) \,.
\end{equation}

 Once can proceed  to evaluate $\psi({x,y,}\Delta t)$, by making the assumption that the action of the unitary operator involving $\hat{H_{0}}$ on $\psi({\bar{x},\bar{y},0})$ gives only a phase factor, which is a valid approximation  for vacuum state and a slightly squeezed vacuum state i.e. with small $r_k$. Hence, we have the following expression for the time evolved wave packet:
\begin{eqnarray}\label{iniwave}
\psi({\bar{x},\bar{y}},\Delta t)&=& {\left(\frac{\Re{(A_{1})}}{\pi}\right)}^{1/4} {\left(\frac{\Re{(A_{2})}}{\pi}\right)}^{1/4} \exp{\left(-\frac{i}{\hbar}\Phi\,{\Delta t} \right)}\nonumber \\
& &\hspace{-2em} \exp{\left({\frac{-i {\Delta t} \mkern1mu} {\hbar}}\hat{V_{\lambda}}\right)} \exp{\left(-\frac{1}{2}({A}_{1}{\bar{x}}^{2}+iB_{1}\bar{x}+iC_{1})\right)} \nonumber \\ 
& & \hspace{-2em} \exp{\left(-\frac{1}{2}({A}_{2}{\bar{y}}^{2}+iB_{2}\bar{y}+iC_{2})\right)}, \nonumber \\
\end{eqnarray}
where $\exp{\left(-i\Phi\,\Delta t/\hbar \right)}$ is the phase factor resulting from the above assumption. Note that, a simple phase factor has no bearing on the results of linear entropy. The interaction terms inside potential $\hat{V}_{\lambda}$ determines the entanglement dynamics. It is already been shown that $\hat{V}_{\lambda}=\lambda \bar{x} \bar{y}$. For the analytical calculation of entanglement entropy, one needs to simplify the expression of $\hat{V}_{\lambda}$ potential.
Author had proved in his previous explorations, the entanglement dynamics in relation to 2D chaotic Hamiltonians with a more general interaction term $x^m\,y^n$\cite{skj_epjd}. Following the analysis in Ref.~\cite{skj_epjd}, one can  derive the results for the entanglement dynamics for any higher order terms appears due to the gravitational interaction but the analysis is limited to short time with weak interaction parameter. In this work, strong interaction parameter regime is explored. For simplicity one can easily ignore the higher order terms, since the separation between the particle is larger compared to the quantum mechanical oscillation. Significant contributions of the interaction term comes from $\frac{\beta}{L^2}\, x\,y$ term by taking $n=2$ and $r=1$. Note that $r=0$ and $r=n$ doesn't contain any interaction terms.  Hence our problem reduces into  finding the entanglement dynamics in a Hamiltonian with a linear interaction term in $\bar{x}$ and $\bar{y}$:
\begin{equation}\label{Linham}
H=\frac{1}{2}(p_{x}^{2}+p_{y}^{2})+\frac{1}{2}k' (\bar{x}^{2}+\bar{y}^{2})+\lambda\, \bar{x} \,\bar{y}.
 \end{equation}
where $\lambda=2\beta/L^2=2GM^2/L^3$.
Time evolving bipartite wavefunction is given by, 
\begin{eqnarray}\label{iniwave2}
& & \psi({\bar{x},\bar{y}},\Delta t) = {\left(\frac{\Re{(A_{1})}}{\pi}\right)}^{1/4} {\left(\frac{\Re{(A_{2})}}{\pi}\right)}^{1/4} \exp{\left(-\frac{i}{\hbar}\Phi\,{\Delta t} \right)}\nonumber \\& & \exp{\left({\frac{-i {\Delta t} \mkern1mu} {\hbar}}\lambda \bar{x} \bar{y}\right)} \exp{\left(-\frac{1}{2}({A}_{1}{\bar{x}}^{2}+iB_{1}\bar{x}+iC_{1})\right)} \nonumber \\ & & \exp{\left(-\frac{1}{2}({A}_{2}{\bar{y}}^{2}+iB_{2}\bar{y}+iC_{2})\right)}. \nonumber \\
\end{eqnarray}
Using Eq.~\ref{ReduDen}, one can find the reduced density matrix of the first subsystem by integrating the second particle state, thus obtains,
\begin{eqnarray}\label{iniwave3}
& & \rho({\bar{x},\bar{z}},\Delta t)= {\left(\frac{\Re{(A_{1})}}{\pi}\right)}^{1/2} {\left(\frac{\Re{(A_{2})}}{\pi}\right)}^{1/2} \nonumber \\
& & \exp{\left(-({A}_{1}{\bar{x}}^{2}+{A}_{1}^{*}{\bar{z}}^{2}+iB_{1}\bar{x}-iB_{1}\bar{z})\right)} \nonumber \\ 
& & \int{\exp{\left(-\frac{1}{2}(({A}_{2}+{A}_{2}^{*}){\bar{y}}^{2}+i\bar{\lambda}(\bar{x}-\bar{z})\bar{y})\right)}\,d\bar{y}}. \nonumber \\
\end{eqnarray}
Defining a new parameter $\bar{\lambda}=\lambda\frac{ {\Delta t} \mkern1mu} {\hbar}$ for convenience, one can rewrite the reduced density matrix in the following manner,

\begin{eqnarray}\label{iniwave4}
& & \rho({\bar{x},\bar{z}},\Delta t)= {\left(\frac{\Re{(A_{1})}}{\pi}\right)}^{1/2} {\left(\frac{\Re{(A_{2})}}{\pi}\right)}^{1/2} \nonumber \\
& &  \exp{\left(-\frac{1}{2}({A}_{1}{\bar{x}}^{2}+{A}_{1}^{*}{\bar{z}}^{2}+iB_{1}\bar{x}-iB_{1}\bar{z})\right)} \nonumber \\ 
& & \int{\exp{\left(-(\Re{({A}_{2})}{\bar{y}}^{2}+i\bar{\lambda}(\bar{x}-\bar{z})\bar{y})\right)}\,d\bar{y}}.\nonumber \\
\end{eqnarray}
Using the well known gaussian integral result, $\int_{-\infty}^{\infty}e^{- (a x^2 + b x + c)}\,dx=\sqrt{\frac{\pi}{a}}\,e^{\frac{b^2}{4a}-c}$, one can find the reduced density matrix $\rho_{1}({\bar{x},\bar{z}},\Delta t)$ and it is given by,
\begin{eqnarray}\label{iniwave5}
& & \rho_{1}({\bar{x},\bar{z}},\Delta t)= {\left(\frac{\Re{(A_{1})}}{\pi}\right)}^{1/2} \nonumber \\
& & \exp{\left(-\frac{1}{2}({A}_{1}{\bar{x}}^{2}+{A}_{1}^{*}{\bar{z}}^{2})-iB_{1}(\bar{x}-\bar{z})\right)} \nonumber \\ 
& &\exp{\left( -\frac{{\bar{\lambda}}^2(\bar{x}-\bar{z})^2}{4\Re{(A_{2})}}\right)}.\nonumber \\
\end{eqnarray}
Note that we have taken  $A_{k}$ as a  complex number ($A_{k} \in \mathbb{C}$) and $B_{k}$ as a real parameter ($B_{k} \in \mathbb{R}$) for $k=1$ and $k=2$.
Now the reduced density matrix $\rho_{1}$ of the first system is obtained, 
\begin{eqnarray}\label{iniwave6}
& & \rho_{1}({\bar{x},\bar{z}},\Delta t)= {\left(\frac{\Re{(A_{1})}}{\pi}\right)}^{1/2} \nonumber \\
& & \exp{\left(-\frac{1}{2}({A}_{1}{\bar{x}}^{2}+{A}_{1}^{*}{\bar{z}}^{2}+iB_{1}(\bar{x}-\bar{z})\right)} \nonumber \\ 
& &\exp{\left( -\frac{{\bar{\lambda}}^2(\bar{x}-\bar{z})^2}{4\Re{(A_{2})}}\right)}.\nonumber \\
\end{eqnarray}
In order to compute the linear entropy of entanglement, one can proceed to evaluate the double integral given in Eq.~\ref{TrSqeq},
\begin{eqnarray}\label{iniwave7}
& & \mbox{Tr}\left[\rho_{1}({\bar{x},\bar{z}},\Delta t)^2\right]= {\left(\frac{\Re{(A_{1})}}{\pi}\right)} \nonumber \\
& &\iint{\exp{\left(-\Re{({A}_{1})}{(\bar{x}^{2}+\bar{z}^{2})}\right)} \exp{\left( -\frac{{\bar{\lambda}}^2(\bar{x}-\bar{z})^2}{2\Re{(A_{2})}}\right)}d\bar{x} d\bar{z}}.\nonumber \\
\end{eqnarray}

Defining a new parameter $\beta^2$ in the following manner will simplify the expression for $\mbox{Tr}\left[\rho({\bar{x},\bar{z}},\Delta t)^2\right]$, i.e,
\begin{equation}
 \beta^2=\frac{\bar{\lambda}^2}{2\Re{(A_{1})}\Re{(A_{2})}}=\frac{\lambda^2 \Delta t^2}{2\hbar^2 \Re{(A_{1})}\Re{(A_{2})}}.
\end{equation}
Representing $\beta^2$ in terms of squeezing parameters, one can get
\begin{equation}\label{betaexpr}
 \beta^2=\frac{\lambda^2 \Delta t^2}{2} \frac{{|\eta_{1}|}^{2} {|\eta_{2}|}^{2}}{\Re{(\eta_{1})}\Re{(\eta_{2})}}
\end{equation}
Rewriting the trace of the square of the reduced density matrix of the first subsystem $ \rho_{1}({\bar{x},\bar{z}},\Delta t)$ in terms of $\beta$ gives,
\begin{eqnarray}\label{Reduden1}
& & \mbox{Tr}\left[\rho_{1}({\bar{x},\bar{z}},\Delta t)^2\right]= {\left(\frac{\Re{(A_{1})}}{\pi}\right)} \nonumber \\
& &\int{\exp{{\left(-\Re{({A}_{1})}(x^2+z^2+\beta^2(x-z)^2 \right)}}d\bar{x} }\nonumber \\
\end{eqnarray}

%

Rearranging the terms and isolate the term to perform the $\bar{z}$ integration first,

\begin{eqnarray}\label{Redudensq}
& & \mbox{Tr}\left[\rho_{1}({\bar{x},\bar{z}},\Delta t)^2\right]= {\left(\frac{\Re{(A_{1})}}{\pi}\right)} \nonumber \\
& &\iint{e^{\left(-\Re{({A}_{1})}(1+\beta^2)\,{\bar{x}^{2}}\right)} e^{\left( -\Re{({A}_{1})}\big((1+\beta^2)\bar{z}^2 -2\beta^2 \bar{x}\bar{z}\big)\right)}d\bar{z}\,d\bar{x} }\nonumber \\
\end{eqnarray}
One can perform the $\bar{z}$ integral via Gaussian integrals and can obtain the following expression, 
\begin{eqnarray}
& & \int{\exp{\left( -\Re{({A}_{1})}\big((1+\beta^2)\bar{z}^2 -2\beta^2 \bar{x}\bar{z}\big)\right)}d\bar{z}}\nonumber \\& &=\sqrt{\frac{\pi}{\Re{({A}_{1})}}}\frac{1}{\sqrt{1+\beta^{2}}} \exp{\left(\frac{-\beta^4}{1+\beta^2}\, {\Re{({A}_{1})}}\,x^2\right)}.\nonumber \\
\end{eqnarray}
Substituting this back into the Eq.~\ref{Redudensq},  one can obtain a pure gaussian integral in $\bar{x}$ as shown,
\begin{eqnarray}\label{iniwave8}
& & \mbox{Tr}\left[\rho_{1}({\bar{x},\bar{z}},\Delta t)^2\right]= \sqrt{\left(\frac{\Re{(A_{1})}}{\pi}\right)} \nonumber \\
& &\int{\exp{\left(-\Re{({A}_{1})}\big(1+\beta^2-\frac{\beta^4}{1+\beta^2}\big)\,{\bar{x}^{2}}\right)} d\bar{x} }.\nonumber \\
\end{eqnarray}
Now, one can easily do the $\bar{x}$ integration since it is a simple standard Gaussian integral and one can obtain the following expression,
\begin{eqnarray}\label{trace_rho_sq}
& & \mbox{Tr}\left[\rho_{1}({\bar{x},\bar{z}},\Delta t)^2\right]= \sqrt{\left(\frac{\Re{(A_{1})}}{\pi}\right)} \frac{1}{\sqrt{1+\beta^{2}}}\nonumber \\
& &\int{\exp{\left(-\Re{({A}_{1})}\big(1+\beta^2-\frac{\beta^4}{1+\beta^2}\big)\,{\bar{x}^{2}}\right)} d\bar{x} }.\nonumber \\
\end{eqnarray}
\begin{eqnarray}\label{iniwave9}
\mbox{Tr}\left[\rho_{1}({\bar{x},\bar{z}},\Delta t)^2\right]=  \frac{1}{\sqrt{1+\beta^{2}}} \frac{1}{\sqrt{\big(1+\beta^2-\frac{\beta^4}{1+\beta^2}\big)}}
\end{eqnarray}

\begin{eqnarray}\label{iniwave10}
\mbox{Tr}\left[\rho_{1}({\bar{x},\bar{z}},\Delta t)^2\right]=  \frac{1}{\sqrt{1+2\beta^{2}}} 
\end{eqnarray}
Now the linear entropy of entanglement can be found to be,
\begin{equation}\label{slfin}
S_{L}(t)=1- \frac{1}{\sqrt{1+2\beta^{2}}}. 
\end{equation}
From the analytical expression, it can be seen that once the interaction strength is zero $\beta=0$, the entanglement entropy correctly goes to zero, since the initial state will become a tensor product state of squeezed coherent states.
\section{Entanglement entropy in weak interaction regime}
Using Eq.~\ref{slfin} and expanding $S_{L}(t)$ for small values of $\beta$, one can obtain the following result
\begin{equation}\label{sl_appx}
S_{L}(t)=\beta^{2}.
\end{equation}

Applying our earlier theoretical analysis\cite{skj_epjd}, the short time entanglement for the squeezed vacuum state under the interaction potential $\lambda \, x^{m} y^{n}$, as a special case ($m=1,n=1$), we found  the expression for linear entropy as follows, 
\begin{eqnarray}\label{Slnfin1}
S_{L}(t)= \frac{1}{2} \, \lambda^2  \, \left(\frac{{|\eta_{1}|}^{2}}{\Re(\eta_{1})}\right) \left(\frac{{|\eta_{2}|}^{2}}{\Re(\eta_{2})}\right){\Delta t}^2 .
 \end{eqnarray}
In our case, expressing $\lambda$ in terms of the Gravitational constant $G$, one can obtain,
\begin{eqnarray}\label{Slnfin}
S_{L}(t)= \frac{1}{2} \, \Bigl(\frac{G^2m^2}{L^6} \Bigr)  \, \left(\frac{{|\eta_{1}|}^{2}}{\Re(\eta_{1})}\right) \left(\frac{{|\eta_{2}|}^{2}}{\Re(\eta_{2})}\right){\Delta t}^2 .
 \end{eqnarray}
Here ${\Re(\eta_{k})}$ denotes the real part of the parameter $\eta_{k}$. It is to be noted that, Eq.~\ref{Slnfin}  describes the short time dynamics of the entanglement entropy.  Inspecting the expression for $\beta$ given in Eq.~\ref{betaexpr}, one can easily identify that we got the same expression as in Eq.~\ref{sl_appx}. Hence in the weak coupling regime our result agrees with the previously found expression~\cite{skj_epjd}.
For real squeezing parameter, this expression reduces to
\begin{eqnarray}\label{kerr_lin_sqz}
S_{L}(t)= \frac{1}{2} \,  \Bigl(\frac{G^2m^2}{L^6} \Bigr) {|\eta_{1}|}{|\eta_{2}|} {\Delta t}^2.
 \end{eqnarray}
It is interesting to note that if a sufficiently higher squeezing strength can be achieved, one can experimentally detect these gravitationally induced entanglements via squeezed coherent state. 

For coherent vacuum state, ${|\eta_{1}|}=1$ and ${|\eta_{2}|}=1$. The linear entropy of entanglement given in Eq.~\ref{kerr_lin_sqz} then further simplifies to
\begin{eqnarray}\label{kerr_lin_coh}
S_{L}(t)= \frac{1}{2} \Bigl(\frac{G^2m^2}{L^6} \Bigr) \, {\Delta t}^2 .
 \end{eqnarray}
One can see that the entanglement entropy increases quadratically with respect to the time.  In the exact solution, this entanglement evolution for a long time will give a different dynamics. This expression valid only for short time for slightly squeezed states due to the assumptions we have taken into arrive at this expression. Hence we will numerically compute the long time entanglement dynamics to have a clear picture on the effect of squeezing.

\section{Covariance matrix verification of the entanglement dynamics}\label{sec:covmat}
The linear entropy of entanglement derived above through explicit Gaussian integrations can be verified in a more direct and elegant manner using the covariance matrix formalism of Gaussian quantum states~\cite{Weedbrook12}. Since the initial squeezed coherent states are Gaussian and the effective interaction Hamiltonian $\hat{V}_{\lambda}=\lambda\,\bar{x}\,\bar{y}$ is quadratic, the state remains Gaussian at all times and is completely characterized by its first moments and its covariance matrix. All the entanglement information is therefore encoded in the covariance matrix, and the first moments (which cannot influence entanglement) may be discarded.

For the phase-space vector $\hat{\bm{\xi}}=(\bar{x},p_{\bar{x}},\bar{y},p_{\bar{y}})^{T}$, the covariance matrix is defined as $\sigma_{ij}=\frac{1}{2}\langle \hat{\xi}_i\hat{\xi}_j+\hat{\xi}_j\hat{\xi}_i\rangle-\langle\hat{\xi}_i\rangle\langle\hat{\xi}_j\rangle$. For the initial product state given in Eq.~\ref{2d_sq_vaccum}, characterized by the complex width parameters $A_{1}$ and $A_{2}$, the single-mode blocks of the initial covariance matrix are
\begin{equation}\label{eq:cm_initial}
\sigma_{k}(0)=\frac{1}{2\Re(A_{k})}
\begin{pmatrix}
1 & -\hbar\,\Im(A_{k})\\[2pt]
-\hbar\,\Im(A_{k}) & \hbar^{2}{|A_{k}|}^{2}
\end{pmatrix}, \quad k=1,2,
\end{equation}
with vanishing cross-correlations between the two modes, so that $\sigma(0)=\sigma_{1}(0)\oplus\sigma_{2}(0)$. One can readily check that $\det\sigma_{k}(0)=\hbar^{2}/4$, as required for a pure Gaussian state.

In the short-time gate approximation of Eq.~\ref{Zassenhaus}, the entangling unitary $\exp{\left(-\frac{i}{\hbar}\lambda\Delta t\,\bar{x}\,\bar{y}\right)}$ acts on the quadratures in the Heisenberg picture as
\begin{equation}
\bar{x}\rightarrow\bar{x}, \quad p_{\bar{x}}\rightarrow p_{\bar{x}}-\lambda\Delta t\,\bar{y}, \quad \bar{y}\rightarrow\bar{y}, \quad p_{\bar{y}}\rightarrow p_{\bar{y}}-\lambda\Delta t\,\bar{x},
\end{equation}
which corresponds to the symplectic transformation $\sigma(\Delta t)=S\,\sigma(0)\,S^T$ with
\begin{equation}\label{eq:symplectic}
S=
\begin{pmatrix}
1 & 0 & 0 & 0\\
0 & 1 & -\lambda\Delta t & 0\\
0 & 0 & 1 & 0\\
-\lambda\Delta t & 0 & 0 & 1
\end{pmatrix}.
\end{equation}
This transformation leaves the position variances untouched and feeds the position fluctuations of each mode into the momentum fluctuations of the other mode. The reduced covariance matrix of the first oscillator (the upper-left $2\times2$ block of $\sigma(\Delta t)$) becomes
\begin{equation}
\sigma_{1}(\Delta t)=\sigma_{1}(0)+
\begin{pmatrix}
0 & 0\\
0 & {(\lambda\Delta t)}^{2}\,\langle\bar{y}^{2}\rangle
\end{pmatrix},
\end{equation}
with $\langle\bar{y}^{2}\rangle=1/(2\Re(A_{2}))$. Its determinant evaluates to
\begin{equation}
\det\sigma_{1}(\Delta t)=\frac{\hbar^{2}}{4}+\frac{{(\lambda\Delta t)}^{2}}{4\,\Re(A_{1})\Re(A_{2})}=\frac{\hbar^{2}}{4}\left(1+2\beta^{2}\right),
\end{equation}
where $\beta^{2}={\bar{\lambda}}^{2}/(2\Re(A_{1})\Re(A_{2}))$ is exactly the parameter introduced in Eq.~\ref{betaexpr}. The purity of a single-mode Gaussian state is given by $\mu_{1}=\mbox{Tr}(\hat{\rho}_{1}^{2})=\hbar/(2\sqrt{\det\sigma_{1}})$, and hence
\begin{equation}\label{eq:cm_purity}
\mu_{1}=\frac{1}{\sqrt{1+2\beta^{2}}}, \qquad S_{L}(t)=1-\frac{1}{\sqrt{1+2\beta^{2}}},
\end{equation}
in exact agreement with Eq.~\ref{slfin} obtained via the wavefunction and Gaussian-integral route. The covariance matrix method thus provides an independent confirmation of our analytical result.

The covariance matrix picture also makes the role of squeezing transparent. For momentum-squeezed vacuum states with real squeezing parameters ($\theta_{k}=0$), one has $A_{k}=({M\omega}/{\hbar})\,e^{-2r_{k}}$, so the position variances are amplified as $\langle\bar{x}^{2}\rangle=({\hbar}/{2M\omega})\,e^{2r_{1}}$ and $\langle\bar{y}^{2}\rangle=({\hbar}/{2M\omega})\,e^{2r_{2}}$. Introducing the dimensionless gate strength $\kappa=\lambda\Delta t/(M\omega)=2GM\Delta t/(\omega L^{3})$, the purity takes the form
\begin{equation}\label{eq:cm_exponential}
\mu_{1}={\left(1+\kappa^{2}\,e^{2(r_{1}+r_{2})}\right)}^{-1/2},
\end{equation}
so that in the weak coupling regime the linear entropy of entanglement is
\begin{equation}\label{eq:cm_SL_exp}
S_{L}(t)\approx\frac{1}{2}\,\kappa^{2}\,e^{2(r_{1}+r_{2})}=\frac{1}{2}{\left(\frac{2GM\Delta t}{\omega L^{3}}\right)}^{2} e^{2(r_{1}+r_{2})}.
\end{equation}
This compact covariance matrix result shows that the gravitationally induced entanglement is exponentially enhanced in the total squeezing $r_{1}+r_{2}$ of the initial momentum-squeezed states. Gravity acts on the two massive oscillators as a continuous-variable entangling gate whose entangling power is amplified exponentially by the initial quantum squeezing. This is fully consistent with the perturbative expression Eq.~\ref{Slnfin}, of which Eq.~\ref{eq:cm_SL_exp} is the momentum-squeezed special case.

\section{Exact Entanglement Dynamics Using Squeezed Coherent State }
\begin{figure}[h]
    \includegraphics[width=7cm,height=4cm]{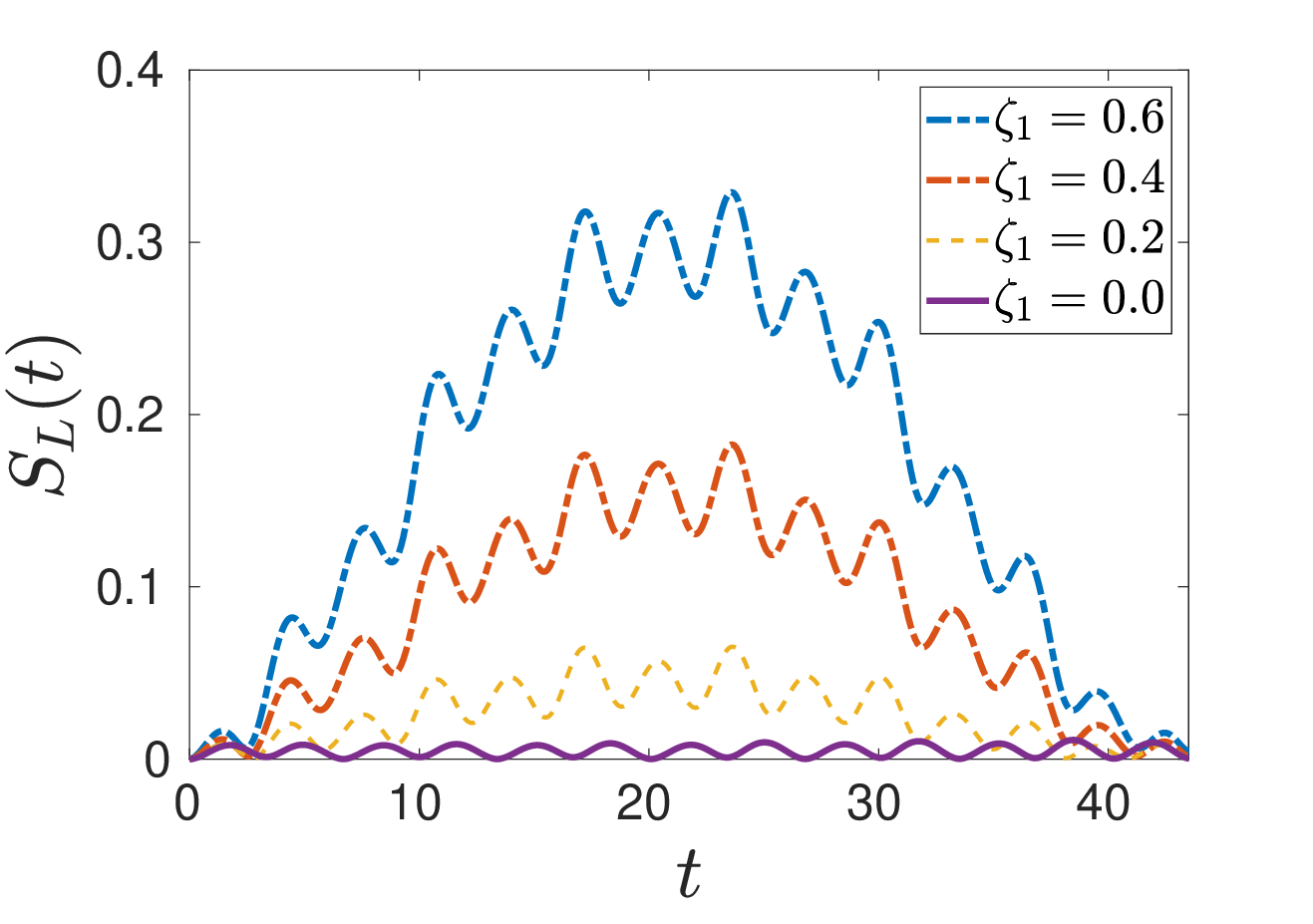}
   \caption{Figure shows the entanglement dynamics of different squeezed vacuum states where only the first oscillator mode is squeezed and the second mode is left in vacuum state. The tensor product state of squeezed vacuum and vacuum state is time evolved using the exact Hamiltonian given in Eq.~\ref{GenHam2}. It can be seen that when $\zeta_{1}=0$, both the state is the tensor product of vacuum state and entanglement dynamics is very small, but periodic in nature (see magenta line). Once we increase the squeezing parameter in steps of $0.2$ upto $\zeta_{1}=0.6$, there is a gradual increase in both initial entanglement as well as the maximum value of the entanglement (see the blue curve). }
   \label{fig:second}
\end{figure}
\begin{figure}[h]
    \includegraphics[width=7cm,height=4cm]{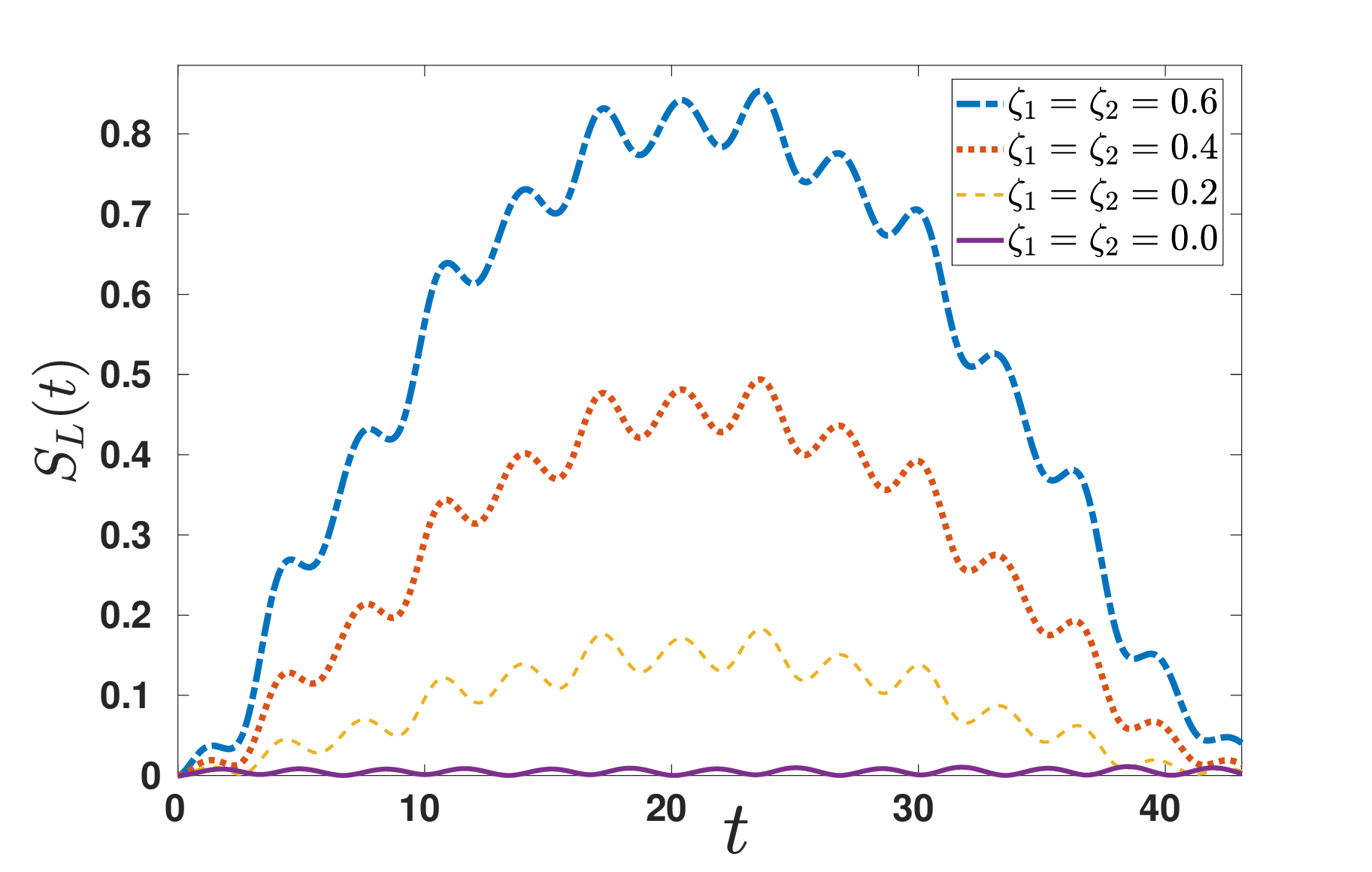}
   \caption{Figure shows the entanglement dynamics of equally squeezed vacuum states $(\alpha_{k}=0)$, where both oscillator modes are equally squeezed $\zeta_{1}=\zeta_{2}$.  The tensor product equally squeezed vacuum state is time evolved using the exact Hamiltonian given in Eq.~\ref{GenHam2}. It can be seen that when $\zeta_{1}=0$, both the state is the tensor product of vacuum state and entanglement dynamics is very small, but periodic in nature (see magenta line). Once we increase the squeezing parameter ($\zeta_{1}=\zeta_{2}$) in steps of $0.2$ upto $\zeta_{1}=\zeta_{2}=0.6$, there is a gradual increase in both initial entanglement as well as the maximum value of the entanglement (see the blue curve). Comparing with Fig.~\ref{fig:second}, it can be seen that equally squeezed state has higher entanglement maxima compared to one oscillator mode squeezed state.}
   \label{fig:third}
\end{figure}
In this section, we will analyze the exact system numerically and we have taken the Hamiltonian given in Eq.~\ref{GenHam2}. As we already know, the short time entanglement entropy will be significantly different from the exact long time dynamics. In order to numerically evaluate the entanglement dynamics exactly, we follow the numerical procedure outlined in Refs.~\cite{Feit_Fleck1,Feit_Fleck2}.  The split operator algorithm provided in Refs.~\cite{Feit_Fleck1,Feit_Fleck2} is memory efficient and extremely fast compared with the matrix method. Due to this reason, their algorithm is utilized to find the time evolution of the initial tensor product squeezed coherent state. In Fig.~\ref{fig:second}, we have evaluated entanglement dynamics of the squeezed vacuum state with several squeezing parameter $\zeta_{1}$ for the first oscillator state, while the second oscillator state $\zeta_{2}$ is kept as zero. We have noticed that changing the squeezing parameters keeping $\zeta_{1}$ the entanglement growth as well as the entanglement maxima is higher. Depending on the strength of squeezing the entanglement is getting enhanced in the system. When  both modes are getting squeezed in the initial tensor product state, the entanglement enhancement is much higher than the single mode squeezed tensor product state. Similarly in Fig.~\ref{fig:third}, we have evaluated entanglement dynamics of the equally squeezed vacuum state with several squeezing parameter $\zeta_{1}=\zeta_{2}$. In equally squeezed state, quantum squeezing is applied on both oscillator modes equally.  We have noticed that changing the squeezing parameter $\zeta_{1}=\zeta_{2}$ the entanglement growth as well as the entanglement maxima is higher compared to the squeezed coherent state with one mode is squeezed. Depending on the strength of squeezing, the gravitationally induced entanglement is getting enhanced in the system.  It can be seen that the analytical expression in the short-time regime can qualitatively speak about this entanglement dependence on the quantum squeezing and it correlates with our numerical results.
\section{Conclusions}
We have shown that Newtonian gravity entangles two harmonically trapped masses and that momentum squeezing amplifies this effect. Within a short-time weak interaction approximation, we derived a closed-form expression for the linear entropy of entanglement, and verified it independently using the covariance matrix formalism of Gaussian states. In the weak-coupling regime the entanglement grows quadratically in time and, for momentum-squeezed initial states and it is exponentially enhanced in the total squeezing $r_{1}+r_{2}$. Here gravity acts as a continuous-variable entangling gate whose entangling power is amplified by the initial momentum squeezing. Exact numerical simulations with the full nonlinear gravitational potential confirm that this enhancement persists beyond the perturbative regime. Squeezed states thus provide a powerful control for bringing gravitationally induced entanglement closer to experimental reach, sharpening the prospects for table-top tests of the quantum nature of gravity.

\bibliographystyle{apsrev4-2}
\bibliography{ref_classical_quantum_all}

\end{document}